\documentstyle[12pt]{article}

\input{psfig}
\def\be{\begin{equation}}
\def\ee{\end{equation}}
\def\bea{\begin{eqnarray}}
\def\eea{\end{eqnarray}}

\begin{document}

\typeout{--- Title page start ---}

\thispagestyle{empty}
\renewcommand{\thefootnote}{\fnsymbol{footnote}}

\begin{tabbing}
\hskip 11.5 cm \= {Imperial/TP/97-98/24}\\
\> February 3 1997 \\
\end{tabbing}


\vskip 1cm
\begin{center}
{\Large\bf {What is the future of causal models of cosmic structure
formation?\footnote{
To appear in the proceedings of the {\em International Workshop on
Particle Physics and the Early Universe} Ambleside, September 1997,
L. Roszkowski {\em Ed.} }  
}} 
\vskip 1.2cm
{\large  Andreas Albrecht}\\
Theory Group, Blackett Laboratory, Imperial College, Prince Consort Road\\
London SW7 2BZ  U.K.\\
\end{center}

\begin{center}
Abstract
\end{center}
Recent research has severely constrained the standard ``defect''
models of cosmic structure formation.  Here I discuss the nature of
the problems with defect models, and place this discussion in the
context of the big picture of cosmic structure formation.
In particular, I classify models of cosmic structure formation as
either ``causal'' or ``acausal'', and ask whether the problems with
the defect models extend to all other causal models.  I argue that
determining the causal nature of the primordial perturbations is
within the reach of modern cosmology, and that such a determination
would yield deep insights into the very early Universe.

\section{Causality and cosmic structure}
There is now overwhelming evidence that the Universe is extremely
homogeneous on large scales.  This fact, combined with the tendency
for gravity to make matter more clumpy as time goes on means that
the early Universe was very smooth indeed.  Still, the early
Universe must have had some very small primordial inhomogeneities in
order to seed the process of gravitational collapse and
produce the observed structure in the Universe.   Because of
the small amplitude required of these initial inhomogeneities, it has
become common in cosmology to speak of the almost perfect homogeneity
and the primordial perturbations as two separate features of the early
Universe.

The popular ``inflationary cosmology'' offers one explanation for
the origin of both these features.  During a period of cosmic inflation
whatever initial homogeneities are present are pushed to such large
scales that they are unobservable.  The fluctuations on observable
scales are predicted based on well-defined calculable processes
which take zero-point quantum fluctuations in the quantum fields and
amplify them into what ultimately become large scale classical
perturbations in the cosmic matter.  If the amplitude of these
perturbations is tuned to be sufficiently small, the inflationary
models predict that the ``Standard Big Bang'' (SBB) epoch which
follows inflation will start out with the required homogeneity and
primordial perturbations.  

Another popular paradigm, typified by the cosmic defect models,
starts with a {\em perfectly} homogeneous universe which is already
experiencing SBB evolution.  At some point, physical processes (such
as a phase transition) then produce inhomogeneities which can seed
cosmic structure.  The origin of the initial perfect homogeneity
might still be inflation (for example some speculate that the fine
tuning problem of the inflationary perturbation amplitude might
actually be solved in nature by producing an absolutely
infinitesimal perturbation amplitude\cite{rome1,alexv}, resulting in an
essentially perfectly homogeneous start to the SBB). Because this
paradigm operates entirely within the SBB, the causality structure
of the SBB is respected.  In 
particular, matter cannot be moved around outside the causal
horizon, and this severely constrains the nature of the perturbations
on large scales.  By contrast, the inflationary models have a very
different causality structure, which allows in principle for arbitrary
adiabatic perturbations to be produced on all relevant scales by the time the
SBB epoch begins.  Models such as inflation, for which
outside-horizon perturbations are present at the start of the SBB
are called, by convention, ``acausal'' models.  Models which start
with a homogeneous SBB and produce the perturbations in accordance
with SBB causality are called ``causal'' models.

Causal and acausal models of cosmic structure offer strikingly
contrasting pictures, and it is thus of quite general interest to
attempt to completely rule out one or the other on the basis of
observations.  In this article I pursue this goal by investigating
the extent to which the known problems with the defect models
reflect more general problems with other causal models.  In the
process I will introduce a very interesting class of causal models (the
``Causal White Noise'' models) which still are just allowed by the
data, but which ultimately will be highly distinguishable from
active models.

\section{The problems with defects}
The defect models have for a long time been considered the primary
alternative to the inflationary origin of the cosmological seeds.
They are classic examples of causal models, in that the SBB is
assumed to start with perfect homogeneity which is then broken by
the formation of defects in a cosmic phase transition.  A ``domain
coarsening'' process then follows in which the dynamics steadily
reduces the number of defects, but at any finite time after the phase
transition there are typically some defects left.  This coarsening
process is usually expected to obey a simple scaling law, in which
the mean defect separation scales linearly with time,
subject to some ``transient'' behavior right after the phase
transition and during the radiation-matter transition.  For some
defect types (strings and textures are favourites) the scaling
property allows the defect energy density to keep at a
constant fraction of the total matter density, and this appears to be
roughly what is required to produce the observed cosmic structure. 

However, there has been a growing understanding that the defect
models have difficulties matching all the current observations.  These
problems came to a head over the past year, as new calculations
appeared which greatly reduced the overall uncertainties in the
predictions from some models\cite{neil}, and showed that the
uncertainties in a
wide class of other models did not help circumvent conflicts with
some particular observations\cite{abr}.   

\begin{figure}
\centerline{\psfig{figure=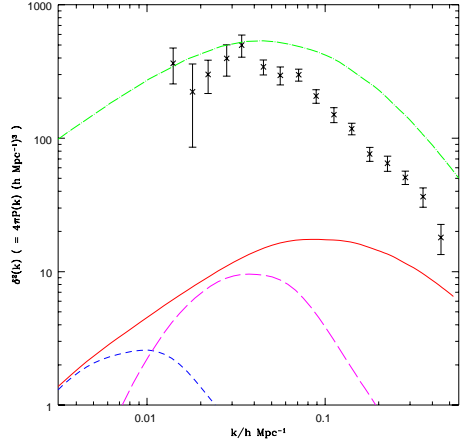,height=2.5in}}
\caption{The power spectrum of the dark matter perturbations for the standard cosmic string model (solid) plotted with
the current observational data, the standard CDM curve (dotted). The
two dashed curves give the partial contributions from two time windows
to either side of $z=100$}
\label{fig1}
\end{figure}

\begin{figure}
\centerline{\psfig{figure=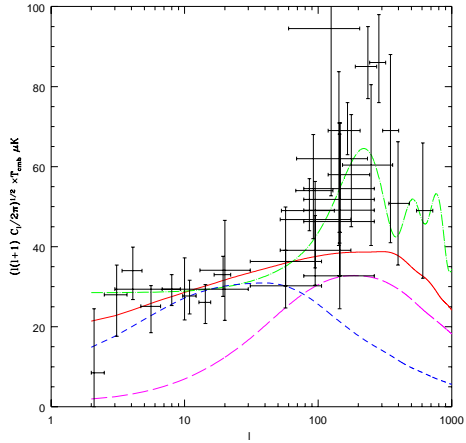,height=2.5in}}
\caption{The (COBE normalized) angular power spectrum of CMB
anisotropies for the
same models and windows shown in Fig. 1, plotted with the data.}
\label{fig2}
\end{figure}

My collaboration\cite{abr} has emphasized that the major source of
conflict with 
the data was the ``$b_{100}$ problem''.  This is illustrated in
Fig. \ref{fig1} where the predictions from a defect model for the
density field power spectrum are compared with the data as presented by
Peacock and Dodds\cite{PD}.  The overall normalization is fixed
by COBE normalizing the Cosmic Microwave Background (CMB) angular
power spectrum which is shown for the same model in Fig. \ref{fig2}.

A number of discrepancies between theory and data can be observed in
these figures, but we have argued\cite{abr2} that by far the most
robust of these problems is the large gap apparent between the
theory and data around $100h^{-1}$ scales in the matter power
spectrum.  (We found\cite{abr}, for example, that it was easy
to exploit uncertainties in the model to boost power in the CMB
anisotropies at high $l$'s where it appears lacking in Fig. \ref{fig2}.)
We chose to express the main difficulty in terms of the ``$b_{100}$
problem''.  The value of $b_{100}$ is the bias required for the
theory and data to match on $100h^{-1}Mpc$ scales, and $b_{100}=5.4$
for the model pictured.  The extent to which there is a $b_{100}$ problem
is the extent to which such large values of the bias are excluded on
those scales.  Most people seem to be convinced that present data
strongly exclude such large values of $b_{100}$ (and favor
$b_{100}\approx 1$), but a minority still feel it is
too early to tell.

The two ``time windows'' illustrated in Figs \ref{fig1} and
\ref{fig2} tell an important part of the story.  The defects produce
perturbations throughout time, and the perturbations on a given
scale are produced predominantly by the defect motions during a
finite period of time.  The ``window'' curves in these two figures
show that the COBE normalization and matter fluctuations which
are relevant to $b_{100}$ are produced during two different time
windows, to either side of the redshift $z=100$.  This means that the
$b_{100}$ problem has a lot to do with how the defect motions in the
two different time windows are related.  In the calculations shown in
Figs \ref{fig1} and \ref{fig2} this relationship was provided by
simply {\em assuming} the standard scaling law.  We have
shown\cite{abr,abr2}
that extreme deviations from  scaling were
required to resolve the $b_{100}$ problem.  But scaling is only one
factor.  When we {\em did} violate scaling sufficiently to solve the
$b_{100}$ problem, we found that CMB power for $l > 100$ was greatly
overproduced.  The problems with defects are connected both with
their scaling properties and their tendency to overproduce CMB
anisotropies relative to density perturbations.  These issues are
discussed in more detail elsewhere\cite{abr2}.

\section{Other causal models}
It is interesting to note that in contrast to the models presented
above, the nearly scale invariant adiabatic spectrum predicted by most
inflation models has no trouble with $b_{100}$.  Contrasting this with
how difficult 
it is for the defect models to reproduce this feature certainly boosts
the standing of the inflation-based models.

To what extent do the problems with standard defect models extend to
other causal models?  Right now there are a number of ``workable'' alternative
causal models. The proposals of Turok\cite{mimic} and
Durrer and Sakellariadou\cite{durrer} consider causal perturbations
which obey standard 
scaling laws, but which are designed
to produce less power in the CMB anisotropies, per matter power, than
the standard defect models.  At the moment these models are 
phenomenological, and do not correspond to a specific microphysical
picture.  Still, I believe they are well worth considering.  There are
a number of ways these models can distinguish themselves from
acausal models, and polarisation is probably a particularly good
discriminator\cite{s-z}.

One can expect, however, that many causal processes will produce power in
the CMB and matter power spectra in similar proportions to the defect
models.  Thus, it is also interesting to investigate how well deviations
from the scaling law can achieve better agreement with the data in
these models.  One such example is the defect models in
non-Einstein-De Sitter cosmologies.  These are expected to exhibit
significant deviations from scaling which might prove sufficient to
produce a viable model\cite{abr3,shellard}.
In the next section I will discuss another type of non-scaling causal model which
I find particularly intriguing: The ``Causal White Noise'' models.

\section{Causal white noise}

The defect models are often called ``active'' models, because they
involve a component of matter (the defects) which evolves in a highly
random 
non-linear manner throughout time, seeding perturbations all along.   By
contrast inflationary models are ``passive'' models, which evolve a
set of initial perturbations in an essentially linear manner until
non-linear gravitational effects set in at late times

One characteristic of causal active perturbations is that at any given
time, they have been unable to move matter around outside the causal
horizon which applies to that process.  While a completely random
process can produce white noise ($P(k) \propto k^0$) on large scales,
the overall mass is constrained to have a $P(k) \propto k^4$ behavior
on large scales for a {\em causal} random processes.

While investigating various non-scaling causal models, we have made
the following remarkable discovery:  A causal white noise spectrum
passes the COBE normalized $b_{100}$ test beautifully.  This means that
{\em any} causal process which has a sufficiently small maximum causal
horizon, and which gets the COBE normalization right, will fit the
galaxy clustering data on $100h^{-1}$Mpc scales.  

\begin{figure}
\centerline{\psfig{figure=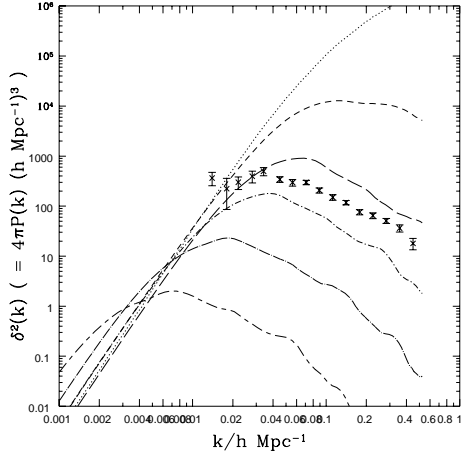,height=2.5in}}
\caption{The (COBE normalized) angular power spectrum of CMB
anisotropies for the
same models and windows shown in Fig. 1, plotted with the data.}
\label{fig3}
\end{figure}

Figure \ref{fig3} illustrates this point.  To produce Fig. \ref{fig3}
we used the same ``standard string'' model which was described in
\cite{abr2} but we turned the string sources on and off (in a causal way)
so that they were only active during a finite time window\cite{abrw}.
Figure \ref{fig3} shows the COBE normalized matter power spectra for
different time windows.  We have found that as long as we turn the
sources off before some critical redshift ($z_c \approx 100$) the
power spectrum passes nicely through the large scale data.  In order
to fit the smaller scale data, more details need to be specified.
Both the specifics of the non-linear processes on scales {\em inside}
the horizon and the type of dark matter are crucial, and we have
already found a number of combinations which provide a good fit.

\begin{figure}
\centerline{\psfig{figure=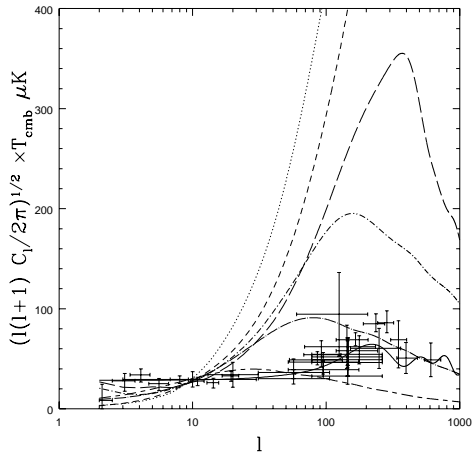,height=2.5in}}
\caption{The (COBE normalized) angular power spectrum of CMB
anisotropies for the
same CWN models depicted in Fig.3 (The upturn exhibited by some curves
at very low $l$ is a numerical artifact.)  The excess power for large
$l$ can be adequately suppressed by a suitable ionization history, and
the small $l$ behavior may produce a better fit in open or $\Lambda$
models.  Standard CDM is also shown.}
\label{fig4}
\end{figure}

However, Fig. \ref{fig4} shows that these Causal White Noise (CWN)
models fare much worse when it comes to the CMB anisotropies.  There
is an excess of power for large $l$ which can be remedied quite well
by choosing a non-standard ionization history\cite{abrw}.  The low $l$
behavior does not look particularly good either, but there is some
hope that low $\Omega_m$ models could produce a better result.

The CWN models are far from being concrete models at this stage.  We
are still working with them at the phenomenological level of trying to
determine what ionization histories and background spacetimes give
these models the best chance of success.
Then there is the matter of finding interesting candidates for the
active sources which produce the noise.  (There have already been
some interesting developments on this front\cite{pt,st}.)

The interesting point at this stage is that CWN models appear to have
what it takes for causal models to avoid a $b_{100}$ problem, and CWN
models reflect very generic processes.  It will be interesting to see if a
concrete viable model can emerge from these ideas.  Again, there
should be ample opportunity to discriminate between CWN models and
acausal models.  One striking difference is the slope of the matter
power spectrum on scales just larger than those on which data
presently exist (Fig. \ref{fig2}).  On these scales the slope of the
power spectrum is much steeper for CWN models than for the standard
scale invariant acausal models.

\section{Conclusions}

The distinction between causal and acausal models of cosmic structure
formation is a very interesting one, which reflects two very different
pictures of the very early Universe.
Recent progress in calculating the predictions from defect models
have brought a wide class of causal models into conflict with the
data.  Some interesting causal models still remain viable, but
there appear to be sufficiently many discriminating features that future
experiments should be able to determine quite clearly the causal
properties of cosmic structure formation.

\section*{Acknowledgements}
I would like to thank R. Battye, J. Robinson and
J. Weller for the enjoyable collaborations which produced this work.
I would also like to thank D. Spergel and N. Bahcall for comments on
the Causal White Noise models. This work is supported by PPARC.


\end{document}